\documentclass[pra,twocolumn,groupedaddress,showpacs,floatfix]{revtex4}

\usepackage[usenames, dvipsnames]{color}
\usepackage{psfrag}
\usepackage{graphicx}
\usepackage{bm}
\usepackage{amsmath}
\usepackage{amsfonts}
\usepackage{amssymb}
\usepackage{latexsym}
\usepackage{cancel}
\usepackage{bbm}
\usepackage{hyperref}

\def\<{\langle}
\def\>{\rangle}
\newcommand{\be}{\begin{equation}}
\newcommand{\ee}{\end{equation}}
\newcommand{\bea}{\begin{eqnarray}}
\newcommand{\eea}{\end{eqnarray}}

\newcommand{\Ket}[1]{|#1\rangle}
\newcommand{\Bra}[1]{\langle#1|}
\newcommand{\Ketbra}[2]{|#1\rangle \langle#2|}
\newcommand{\Braket}[1]{\langle #1 \rangle}

\begin{document}

\title{Macroscopic quantum jumps and entangled state preparation}
\author{Jeremy Metz$^{1}$ and Almut Beige$^{2}$}
\affiliation{$^1$Blackett Laboratory, Imperial College London, Prince Consort Road, London SW7 2BZ, United Kingdom \\
$^2$The School of Physics and Astronomy, University of Leeds, Leeds LS2 9JT, United Kingdom}

\date{\today}

\begin{abstract}
Recently we predicted a random blinking, i.e.~{\em macroscopic quantum jumps}, in the fluorescence of a laser-driven atom-cavity system [Metz {\em et al.},~Phys.~Rev.~Lett.~{\bf 97}, 040503 (2006)]. Here we analyse the dynamics underlying this effect in detail and show its robustness against parameter fluctuations. Whenever the fluorescence of the system stops, a macroscopic dark period occurs and the atoms are shelved in a maximally entangled ground state. The described setup can therefore be used for the controlled generation of entanglement. Finite photon detector efficiencies do not affect the success rate of the state preparation, which is triggered upon the observation of a macroscopic fluorescence signal. High fidelities can be achieved even in the vicinity of the bad cavity limit due to the inherent role of dissipation in the jump process. 
\end{abstract}

\maketitle

\section{Introduction} \label{sec:intro}

A crucial part of the debate on the foundations of quantum mechanics and its implications for single systems was the existence of quantum jumps \cite{Blatt}. A prominent example is the discussion between Schr\"odinger and Bohr. Schr\"odinger asserted that the application of quantum mechanics to single quantum systems would necessarily lead to nonsense such as quantum jumps. In response Bohr argued that the problem lay with the physics experiments of the time, which he believed unsuitable for the demonstration of their existence \cite{Bohr}. Later, in 1975 Dehmelt pointed out that quantum jumps might occur in the form of {\em macroscopic quantum jumps}, when driving a single three-level atom with appropriate laser fields \cite{shelving}. These manifest themselves as a random telegraph fluorescence signal with periods of constant fluorescence (light periods) interrupted by periods of no fluorescence (dark periods). With the development of ion trapping technology, it indeed became possible to confirm Dehmelt's predictions experimentally \cite{Toschek,Toschek2,Toschek3}.

Theoretical models have been developed to describe macroscopic quantum jumps qualitatively and quantitatively \cite{javanainen,knight,Cook,shelving2}. For example, Ref.~\cite{shelving2} is based on the quantum jump approach \cite{review,Hegerfeldt}, which enables the prediction of all the possible quantum trajectories of a single trapped ion undergoing photon emissions. The mean durations of the macroscopic light and dark periods have been calculated using this approach and were found to be in good agreement with experimental findings. Macroscopic quantum jump experiments have also been performed with and analysed for setups containing not only one but several atoms \cite{exp,BeHe99,HaHe}. In such experiments the number of atoms emitting photons is always discrete, thereby causing random variation between distinct fluorescence levels.

\begin{figure}
\begin{minipage}{\columnwidth}
\begin{center}
\resizebox{\columnwidth}{!}{\rotatebox{0}{\includegraphics {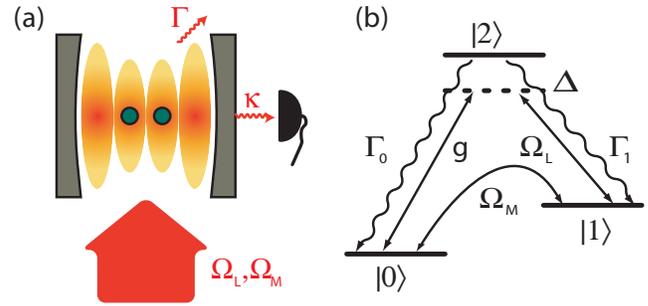}}}
\end{center}
\vspace*{-0.5cm}
\caption{(Colour online) (a) Experimental setup containing two atoms trapped inside an optical cavity and driven by appropriate laser fields. A detector observes the fluorescence leaking out through the cavity mirrors. (b) Level scheme of one of the atoms in the cavity. (c) Macroscopic quantum jumps as they might be recorded by the photon detector.} \label{fig:setup}
\end{minipage}
\end{figure}

In this paper we analyse a system consisting of two laser driven three-level atoms trapped inside an optical cavity as shown in Fig.~\ref{fig:setup}(a). That the trapping of atoms inside such a resonator is experimentally feasible has already been shown by several groups. First experiments have been performed combining atom or ion trapping technology \cite{Chapman2,Meschede2,Kimble,Meschede0,Blattt,Meschede,Meschede3,Meschede4,rempe-cavs} with optical cavities. Meschede's group in Bonn have succeeded in constructing an atomic conveyor belt, which allows one to localise atoms with very high precision \cite{Meschede} and can be combined with an optical cavity. Related experiments are currently carried out in Rempe's group in Garching \cite{rempe-cavs,Meschede2}. Relatively strong atom-cavity couplings have already been achieved in optical resonators mounted on atom chips \cite{Meschede3,Meschede4}. 

The level structure of the atoms considered here is shown in Fig.~\ref{fig:setup}(b). In a recent paper~\cite{MTB-PRL}, we predicted macroscopic quantum jumps in the leakage of the photons through the cavity mirrors of this setup. They occur when the trapping of the atoms and the directions of the incoming laser fields are such that both atoms experience the {\em same} coupling constants. In the following, we denote the coupling strength of the 0--2 transition of each atom to the cavity field by $g$, the laser Rabi frequencies for the 0--1 and the 1--2 transitions by $\Omega_{\rm M}$ and $\Omega_{\rm L}$, and assume
\begin{equation} \label{eqn:conditions}
\Omega_{\rm M} < g , \, \kappa , \, \Gamma  , \, \Omega_{\rm L} \ll \Delta \, .
\end{equation}
Here, $\Gamma$ is the spontaneous decay rate of level 2 and $\kappa$ is the photon leakage rate \cite{kappa}. $\Omega_{\rm M}$ can be realised by using a microwave or a two-photon Raman transition involving level 2 or a fourth level, when direct excitation of the 0--1 transition is not possible. 

The occurrence of macroscopic light and dark periods in the fluorescence of a single ion is also known as {\em electron shelving} \cite{shelving}, since the state of the system remains restricted onto a certain subspace of states within each fluorescence period. In the combined atom-cavity system considered here, the atoms are shelved into the maximally entangled ground state 
\begin{eqnarray} \label{a01}
|a_{01} \rangle &\equiv & (|01 \rangle - |10 \rangle)/\sqrt{2} 
\end{eqnarray}
within each dark period. The described experiment can therefore be used to prepare maximally entangled qubit pairs. The successful state preparation is indicated by the sudden absence of fluorescence. To avoid a return of the system into a light period, the applied laser fields should be turned off when this occurs.

Currently, the practical implementation of quantum computing in atom-cavity systems is limited by the presence of relatively large spontaneous decay rates \cite{JMO}. Many proposed schemes aim at the controlled generation of entanglement through the induction of a coherent time evolution  \cite{Pellizzari1995,Marr2003,ZhengGuo,You2003}. Other proposals employ measurements and dissipation \cite{letter,PachosWalther,Barrett}. However, relatively large decay rates can only be tolerated when operating atom-cavity systems as single photon sources and when generating entanglement via the detection of single photons \cite{Cabrillo,Cabrillo2,Plenio,Lim,Duan-Kimble}. Unfortunately, the scalability of such schemes suffers greatly due to finite photon detector efficiencies. However, when using the observation of macroscopic quantum jumps to trigger the preparation of entangled states, as we describe here, finite detector efficiencies $\eta < 1$ no longer hinder the generation of entanglement. 

The proposed scheme is relatively robust against parameter fluctuations. As we see below, the coupling constants $g$ of the two atoms can vary more than $30 \, \%$ without decreasing the fidelity of the prepared state by more than a few percent. This is due to the postselective nature of the state preparation. As long as the Rabi frequency $\Omega_{\rm M}$ remains the same for both atoms, the antisymmetric state (\ref{a01}) is the only one that does not experience any laser driving. It remains the state with the lowest cavity photon emission rate. The observation of no photons for a relatively long time, therefore indicates that the system is very likely in the maximally entangled state $|a_{01} \rangle$. However, when the $g$'s of the two atoms differ by too much, the probability for the observation of a dark period decreases rapidly. 

Furthermore, our scheme is based on the very dissipation channels that other proposals try to avoid. Cavity decay with the spontaneous photon leakage rate $\kappa$ is responsible for the detector signal within a light period. Spontaneous emission of excited atomic states with decay rate $\Gamma$ is less welcome but plays a crucial role in activating transitions between light and dark periods. Achieving high fidelities is thus possible even in the presence of non-negligible decay rates. We will see that the achievable quality of the prepared state depends primarily on the single atom-cooperativity parameter $C$, defined as
\begin{eqnarray} \label{C}
C &\equiv & \frac{g^2}{\kappa \Gamma} \, .
\end{eqnarray}
For $C \ge1$ and $\eta = 1$ it is possible to achieve fidelities above $0.86$. However, smaller detector efficiencies require larger $C$'s and $\eta C$ becomes the crucial parameter, which determines the achievable fidelities. For example for $\eta = 0.2$, fidelities above $0.9$ require $C \ge 10$. 

There are four sections in this paper. In the following section we examine the phenomenon of macroscopic quantum jumps using a simple toy model as an example. In Section \ref{entangle1} we show that the atom-cavity system in Fig.~\ref{fig:setup} can effectively be reduced to the four-level toy model considered in Section \ref{sec:QJ}. It exhibits macroscopic quantum jumps in an analogous way and we determine the charactristic time scales of the system. In Section \ref{entangle} we outline the creation of entangled pairs of atoms with unit efficiency and calculate the corresponding fidelities. Finally, we summarise our findings in Section \ref{sec:conc}.

\section{Macroscopic quantum jumps in a four-level toy model} \label{sec:QJ}

\begin{figure}
\begin{minipage}{\columnwidth}
\begin{center}
\resizebox{\columnwidth}{!}{\rotatebox{0}{\includegraphics {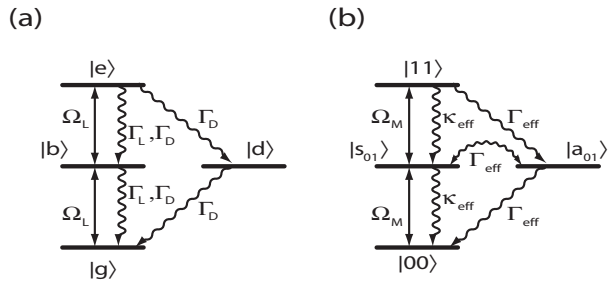}}}
\end{center}
\vspace*{-0.5cm}
\caption{(a) Four-level toy model. We assume that the $g$--$b$ and the $b$--$e$ transition are excited by a resonant laser field. Spontaneous photon emissions occur either with a rate $\Gamma_{\rm L}$ or $\Gamma_{\rm D}$. (b) Effective level scheme of the atom-cavity system shown in Fig.~\ref{fig:setup}, illustrating the effect of the conditional Hamiltonian (\ref{cond4}) and the reset operators (\ref{RR01})-(\ref{RRcav}).} \label{fig:toy model}
\end{minipage}
\end{figure}

In order to gain a qualitative understanding of the phenomenon of macroscopic light and dark periods we now use the quantum jump approach \cite{Hegerfeldt,review} to analyse a four-level {\em toy model}. It's level configuration is shown in Fig.~\ref{fig:toy model}(a). The reason for considering this particular system is that it has great similarities to the effective atom-cavity level scheme in Fig.~\ref{fig:toy model}(b), which we consider in Section \ref{entangle1} for the creation of maximally entangled atom pairs. 

\subsection{Theoretical model} \label{IIA}

The level scheme in Fig.~\ref{fig:toy model}(a) shows two different spontaneous decay channels. If the excited states $|b \>$, $|d \>$ or $|e \>$ are populated, a photon can be emitted with a decay rate $\Gamma_{\rm D}$. Thereby the state $|\psi \rangle$ of the system changes into $R_{\rm D} \, |\psi \rangle/\| \, R_{\rm D} \, |\psi \rangle \, \|$ with the {\em reset} (or jump) operator 
\begin{eqnarray} \label{eqn:toyReset1}
R_{\rm D} &\equiv & \sqrt{\Gamma_{\rm D}} \, \big[ \, \Ketbra{d}{e} + \Ketbra{g}{d} + \Ketbra{b}{e} + \Ketbra{g}{b} \, \big] \, . 
\end{eqnarray}
We also assume that population in $|b \>$ and $|e \>$ can cause another type of photon to be emitted with decay rate $\Gamma_{\rm L}$. In this case, the state vector changes into $R_{\rm L} \, |\psi \rangle/\| \, R_{\rm L} \, |\psi \rangle \, \|$ with the reset operator
\begin{eqnarray} \label{eqn:toyReset2}
R_{\rm L} &\equiv &  \sqrt{\Gamma_{\rm L}} \, \big[ \, |b \> \< e|  + |g \> \< b| \, \big] \, .
\end{eqnarray}
The normalisation of the above reset operators has been chosen such that
\begin{eqnarray} \label{eqn:density}
w_i (\psi) &=& \| \, R_i \, |\psi \rangle \, \|^2  
\end{eqnarray}
is the probability density for an emission given the state $|\psi \rangle$ of the system prior to the emission. Here $i$ equals ${\rm D}$ or ${\rm L}$. In the following we assume that an emission is either of type ${\rm D}$ with decay rate $\Gamma_{\rm D}$ or of type ${\rm L}$ with decay rate $\Gamma_{\rm L}$.

Also shown in Fig.~\ref{fig:toy model}(a) is a laser field, which drives the $g$--$b$ and the $b$--$e$ transition  with Rabi frequency $\Omega_{\rm L}$. The state of the system therefore evolves under the condition of {\em no} photon emission in $(0,t)$ into
\begin{eqnarray} \label{eqn:toyEvolution}
| \psi^0(t) \rangle &=& U_{\rm cond}(t,0) \, |\psi_0 \rangle /\| \, U_{\rm cond}(t,0) \, |\psi_0 \rangle \,  \| 
\end{eqnarray}
with the  conditional Hamiltonian
\begin{eqnarray} \label{eqn:toyHcond}
H_{\rm cond} &=& {\textstyle \frac{1}{2}} \hbar \Omega_{\rm L} \big[ \, \Ketbra{b}{e}+\Ketbra{g}{b} + {\rm H.c.} \, \big] \nonumber\\
&& - {\textstyle \frac{{\rm i}} {2}} \hbar \Gamma_{\rm D} \big[ \, \Ketbra{b}{b} + \Ketbra{d}{d} + 2 \, \Ketbra{e}{e} \, \big] \nonumber\\
&& - {\textstyle {\frac{\rm i}{ 2}}} \hbar \Gamma_{\rm L} \big[ \, \Ketbra{b}{b} + \Ketbra{e}{e} \, \big] \, ,
\end{eqnarray}
if $|\psi_0 \rangle$ is the state of the system at $t=0$. The Hamiltonian (\ref{eqn:toyHcond}) is non-Hermitian. The non-Hermitian terms decrease the relative amount of population in excited states that can cause an emission. This reflects the fact that an observer, who cannot see any photons, learns gradually that the system is more likely to be in a state which cannot emit. 

As an alternative to the quantum jump approach, the four-level toy model in Fig.~\ref{fig:toy model}(a) can be described by the master equation 
\begin{eqnarray} \label{eqn:master}
\dot \rho &=& - {\frac{\rm i} {\hbar}} \big( H_{\rm cond} \, \rho - \rho \, H_{\rm cond}^\dag \big) +  \cal{R}(\rho)
\end{eqnarray}
with 
\begin{equation} \label{eqn:toyReset}
{\cal R}(\rho) = R_{\rm D} \, \rho \, R_{\rm D}^{\dagger} + R_{\rm L} \, \rho \, R_{\rm L}^{\dagger} \, .
\end{equation}
This differential equation predicts the time evolution of the system averaged over an {\em ensemble} of single realisations. Eq.~(\ref{eqn:master}) is therefore a convenient tool for the calculation of unconditioned probabilities, i.e.~probability densities for certain events to take place without prior knowledge about the systems evolution. 

\subsection{Macroscopic light and dark periods} \label{sec:IIB}

For the system considered here, we expect the occurrence of macroscopic quantum jumps when   
\begin{eqnarray} \label{eqn:toy_conditions}
\Gamma_{\rm D} &\ll & \Omega_{\rm L} \, , \Gamma_{\rm L} \, ,
\end{eqnarray}
while $\Omega_{\rm L}$ should be of comparable size to $\Gamma_{\rm L}$ or larger. This ensures that there are two very distinct time scales in the system. To examine the origin of the expected macroscopic light and dark periods, we assume that a photon has just been emitted, thereby triggering reset operation $R_{\rm D}$. Such a quantum jump leaves the system in Fig.~\ref{fig:toy model}(a) in a superposition of the states $|b\>$, $|d\>$ and $|g\>$. Now there are two possible types of dynamics that can occur: 

\begin{figure}
\begin{minipage}{\columnwidth}
\begin{center}
\resizebox{\columnwidth}{!}{\rotatebox{0}{\includegraphics {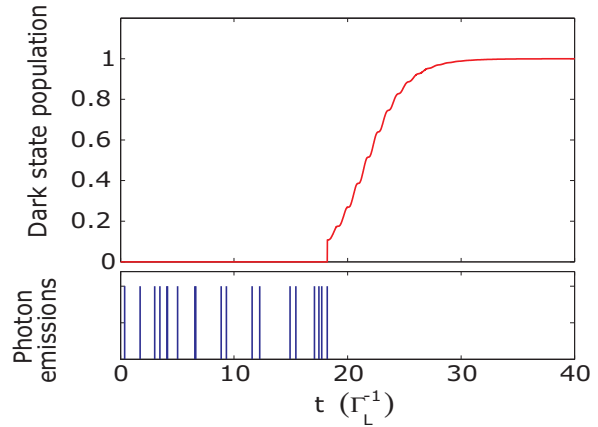}}}
\end{center}
\vspace*{-0.5cm}
\caption{(Colour online) Possible trajectory of the four-level toy model illustrating a transition from a light into a dark period 
obtained from a quantum jump simulation using Eqs.~(\ref{eqn:toyReset1})-(\ref{eqn:toyHcond}) and assuming $\Omega_{\rm L} = \Gamma_{\rm L}$ and $\Gamma_{\rm D} = 10^{-3} \, \Gamma_{\rm L}$. The upper half of the figure shows the population in the dark state $|b \>$ as a function of time, while the vertical lines below mark photon emission times.} \label{fig:toydarkjumps}
\end{minipage}
\end{figure}

\begin{enumerate}
\item Suppose there is a non-zero population in $|d \rangle$ and {\em no} photon emission occurs for a time, which is relatively long compared to $1/\Omega_{\rm L}$ and $1/\Gamma_{\rm L}$. Then the conditional time evolution (\ref{eqn:toyEvolution}) damps away any population in the states $|g \>$ and the excited states $|b \>$ and $|e \>$ despite the applied laser driving. The reason for this is that the normalisation of the state vector (\ref{eqn:toyEvolution}) of the system constantly increases the relative population in the state $|d \rangle$, whose decay rate $\Gamma_{\rm D}$ is relatively low. As illustrated by Fig.~\ref{fig:toydarkjumps}, the no-photon time evolution eventually prepares the system in $|d \>$ with very high fidelity. This state is known as the dark state of the system, since the probability density for a photon emission in this state, $\Gamma_{\rm D}$, is relatively low. The system has entered a {\em macroscopic dark period}. 

Eventually a photon emission will lead to another quantum jump, thereby transferring the system into $\Ket{g}$. At that point, all population in $|d \rangle$ is lost. The system evolves again much more quickly and photons can be emitted at a relatively high rate.

\begin{figure}[t]
\begin{minipage}{\columnwidth}
\begin{center}
\resizebox{\columnwidth}{!}{\rotatebox{0}{\includegraphics{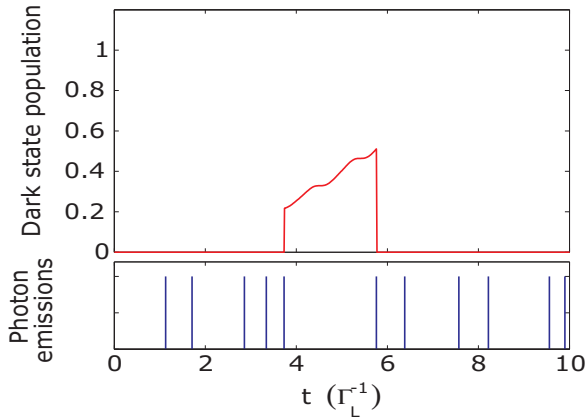}}}
\end{center}
\vspace*{-0.5cm}
\caption{(Colour online) Possible trajectory of the four-level toy model obtained as in Fig.~\ref{fig:toydarkjumps}. Again, a photon emission creates a non-negligible dark state population. However, now another photon is emitted before the dark state population reaches unity and the system remains in a macroscopic light period.} \label{fig:toylightjumps}
\end{minipage}
\end{figure}

\item Alternatively to the above case, another photon emission might occur after a relatively short time, inducing a quantum jump either according to Eq.~(\ref{eqn:toyReset1}) or according to Eq.~(\ref{eqn:toyReset2}). This prepares the system in $|g \>$ or a superposition of $|g \>$ and $|b \>$ and results in a relatively large probability density for subsequent photon emissions. The reason for this is the presence of the relatively strong driving field with Rabi frequency $\Omega_{\rm L}$, which continuously excites the states $|b \>$ and $|e \>$. From there photons can be emitted with the relatively large decay rate $\Gamma_{\rm L}$. Consequently,  the system experiences a {\em macroscopic light period}.

Although the probability density for this is relatively low, a photon emission with decay rate $\Gamma_{\rm D}$ will occur from time to time. This is accompanied by a quantum jump as described by Eq.~(\ref{eqn:toyReset1}) and results in general in the build up of a non-negligible dark state population. In many cases, such an emission is followed by another photon emission relatively shortly afterwards, as shown in Fig.~\ref{fig:toylightjumps}. But eventually, these jumps will result in a transition into a macroscopic dark period and the light period will end.
\end{enumerate}

\subsection{Characteristic time scales} \label{sec:ts1}

We now calculate the mean length of the light and dark periods, $T_{\rm L}$ and $T_{\rm D}$, and the mean time between two photon emissions within a light period, $T_{\rm E}$, analytically. Let us first assume that the system is in a dark period. Ignoring the initial relatively short transition time shown in Fig.~\ref{fig:toydarkjumps}, we can assume that the system is in this case constantly in its dark state $|d \>$. The probability density for leaving this state is thus given by the decay rate $\Gamma_{\rm D}$ at all times. Consequently, the mean length of a dark period equals 
\begin{eqnarray} \label{eqn:Tdarktoy}
T_{\rm D} &=& \frac{1}{\Gamma_{\rm D}} 
\end{eqnarray}  
to a very good approximation. 

To calculate $T_{\rm E}$ and $T_{\rm L}$, we need to know the average state of the system within a light period. To determine this, we note that it equals the steady state $\rho_{\rm ss}$ of the three-level system consisting of the states $|g \>$, $|b \>$ and $|e \>$ given $\Gamma_{\rm D} = 0$. Assuming $\Gamma_{\rm D} = 0$, setting $\dot{\rho}$ in Eq.~(\ref{eqn:master}) equal to zero and using the notation
\begin{eqnarray} \label{xtoy}
x &\equiv & \frac{\Omega_{\rm L}}{\Gamma_{\rm L}} \, ,
\end{eqnarray}
we obtain the steady state populations
\begin{eqnarray}  \label{eqn:toyss}
\langle g| \rho_{\rm ss} |g \rangle &=& \frac {1+x^2+x^4}{1+2\,x^2+3\,x^4} \, , \nonumber\\
\langle b| \rho_{\rm ss} |b \rangle &=& \frac {x^2+x^4}{1+2\,x^2+3\,x^4} \, , \nonumber\\
\langle e| \rho_{\rm ss} |e \rangle &=& \frac{x^4}{1+2\,x^2+3\,x^4} \, . 
\end{eqnarray}
The probability density for a photon emission with $\Gamma_{\rm L}$ within a light period is $\Gamma_{\rm L} (\langle b| \rho_{\rm ss} |b \rangle  + \langle e| \rho_{\rm ss} |e \rangle)$. Its inverse,
\begin{eqnarray} \label{eqn:Temtoy}
T_{\rm E} &=& \frac{1+2\,x^2+3\,x^4}{x^2 + 2 x^4} \cdot \frac{1}{\Gamma_{\rm L}} \, ,
\end{eqnarray}  
equals the mean time between two photons within a light period. 

\begin{figure}[t]
\begin{minipage}{\columnwidth}
\begin{center}
\resizebox{\columnwidth}{!}{\rotatebox{0}{\includegraphics {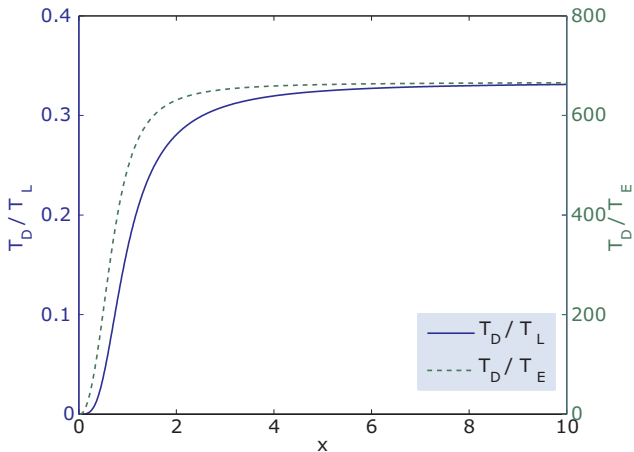}}}
\end{center}
\vspace*{-0.5cm}
\caption{(Colour online) Comparison of the mean length of a dark period $T_{\rm D}$ with the mean length of a light period $T_{\rm L}$ and the mean time $T_{\rm E}$ between photon emissions within a light period as a function of $x$ (c.f.~Eq.~(\ref{xtoy})) for $\Gamma_{\rm D} = 10^{-3} \, \Gamma_{\rm L}$.} \label{fig:toyratios}
\end{minipage}
\end{figure}

The only way to induce a transition into a dark period is a photon emission from the excited state $|e \rangle$ with decay rate $2 \Gamma_{\rm D}$. Immediately after such an emission, the system is in a superposition of $|g \rangle$, $|d \rangle$, and $|b \rangle$. However, in half of the cases such an emission is followed by a no-photon time evolution, which projects the system into the dark state $|d \rangle$. The probability density for an emission with $2 \Gamma_{\rm D}$ multiplied with $\frac{1}{2}$ equals $\Gamma_{\rm D} \langle e| \rho_{\rm ss} |e \rangle$. Its inverse, 
\begin{eqnarray} \label{eqn:Tlighttoy}
T_{\rm L} &=& \frac{1+2\,x^2+3\,x^4}{x^4} \cdot \frac{1}{\Gamma_{\rm D}} \, ,
\end{eqnarray}  
is the mean length of a light period.

In order to ensure the frequent occurrence of dark periods, it is important that $T_{\rm D}$ is not orders of magnitude smaller than $T_{\rm L}$. Otherwise, periods of no fluorescence become very rare. In addition, $T_{\rm D}$ should be much larger than $T_{\rm E}$. Then it is easy to distinguish a dark period from a light period and the detection of no photon for a time large compared to $T_{\rm E}$ indicates the shelving of the system in $|d \>$. From the equations
\begin{eqnarray} \label{ratios}
\frac{T_{\rm D}}{T_{\rm L}} &=& \frac{x^4}{1+2\,x^2+3\,x^4} \, , \nonumber \\
\frac{T_{\rm D}}{T_{\rm E}} &=& \frac{x^2+ 2 x^4}{1+2\,x^2+3\,x^4} \cdot \frac{\Gamma_{\rm L}}{\Gamma_{\rm D}}
\end{eqnarray}  
and Fig.~\ref{fig:toyratios} we see that this requires $\Gamma_{\rm D} \ll \Gamma_{\rm L}$, while $\Omega_{\rm L}$ should be of similar size to $\Gamma_{\rm L}$ or larger, as assumed in the beginning of Section \ref{sec:IIB}. 

\section{Macroscopic quantum jumps in atom-cavity systems} \label{entangle1}

In the previous section we saw that a {\em single} quantum system can be driven such that it produces two distinct periods of fluorescence. The observation of a certain fluorescence level gives us information about the system and corresponds to the shelving of the system in a certain subspace of states. However, macroscopic quantum jumps occur also in {\em composite} quantum systems. An example is the fluorescence from a trap containing more than one atom \cite{exp,BeHe99,HaHe}. In this section, we consider a system consisting of two atoms placed into an optical cavity as shown in Fig.~\ref{fig:setup}(a), each with a level structure as shown in Fig.~\ref{fig:setup}(b). The trapping of the particles and the directions of the incoming laser fields should be such that both atoms experience the same coupling constants. 

\subsection{Theoretical Model} \label{sec:eff}

As in Section \ref{sec:QJ}, we use the quantum jump approach \cite{Hegerfeldt,review} and the master equation to describe the time evolution of this system. The conditional Hamiltonian, which describes its no-photon time evolution, in the interaction picture and within the rotating wave approximation, now equals
\begin{eqnarray} \label{eqn:Hcond1}
H_{\rm cond} &=& \sum_{i=1}^{2} \big[ \, {\textstyle \frac{1}{2}} \hbar \Omega_{\rm L} \, \Ket{1}_{ii}\Bra{2} + {\textstyle {\frac{1}{2}}} \hbar \Omega_{\rm M} \, \Ket{0}_{ii}\Bra{1} + {\rm H.c.} \, \big] \nonumber\\
&& + \sum_{i=1}^{2} \big[ \, \hbar g \, \Ket{0}_{ii}\Bra{2} \, b^\dagger + {\rm H.c.} \, \big] \nonumber \\ 
&& + \hbar  \big( \Delta - {\textstyle {\frac{\rm i}{2}}} \Gamma \big) \, |2 \rangle_{ii} \langle 2| - {\textstyle {\frac{\rm i}{2}}} \hbar \kappa \, b^\dagger b \, .
\end{eqnarray}
Here, $b$ is the annihilation operator for a single photon in the cavity field. Moreover, there are three different distinguishable types of emission. An emission might occur via an atomic decay of the state $|2 \>$ either into $|0 \>$ with rate $\Gamma_0$ or into $|1 \>$ with rate $\Gamma_1$, where 
\begin{eqnarray}
\Gamma_0 + \Gamma_1 = \Gamma \, . 
\end{eqnarray}
In addition, there is the possibility of the leakage of photons through the cavity mirrors with rate $\kappa$. An emission via the 2--$j$ transition changes the density matrix $\rho$ of the system into
\begin{eqnarray} \label{eqn:Rj}
{\cal R}_j (\rho) &=& \sum_{i=1,2} \Gamma_j \, |j \>_{ii} \< 2| \, \rho \, |2 \>_{ii} \< j| \, ,
\end{eqnarray}
when the 0--2 and the 1--2 transition can be distinguished easily due to different polarisations or frequencies of the emitted photons. The leakage of a photon through the cavity mirrors on the other hand changes the state of the system from $\rho$ into 
\begin{eqnarray} \label{eqn:Rcav}
{\cal R}_{\rm C} (\rho) &=& \kappa \, b \, \rho \, b^\dagger \, . 
\end{eqnarray}
The normalisation of the reset states is chosen such that ${\rm Tr} ({\cal R}_j (\rho))$ and ${\rm Tr} ({\cal R}_{\rm C} (\rho))$ are the corresponding emission probabilities. The master equation 
\begin{eqnarray} \label{eqn:master2}
\dot \rho &=& - {\frac{\rm i} {\hbar}} \big( \, H_{\rm cond} \, \rho - \rho \, H_{\rm cond}^\dagger \, \big) +  \sum_{j=0,1} {\cal R}_j(\rho) + \cal{R}_{\rm C}(\rho) \nonumber \\
\end{eqnarray}
is consistent with the quantum jump approach above and describes ensemble-averages.

As the operators (\ref{eqn:Hcond1}), (\ref{eqn:Rj}) and (\ref{eqn:Rcav}) do not distinguish between atom 1 and atom 2, it is convenient to introduce the symmetric and antisymmetric states
\begin{eqnarray} \label{ajk}
|a_{jk} \rangle &\equiv & (|jk \rangle - |kj \rangle)/\sqrt{2} \, , \nonumber \\
|s_{jk} \rangle &\equiv & (|jk \rangle + |kj \rangle)/\sqrt{2} \, .
\end{eqnarray}
Using this notation, Eq.~(\ref{eqn:Hcond1}) becomes
\begin{widetext}
\begin{eqnarray} \label{Hcond2}
H_{\rm cond} &=& {\textstyle \frac{1}{2}} \hbar \Omega_{\rm L} \big[ \Ket{s_{01}}\Bra{s_{02}} +  \Ket{a_{01}}\Bra{a_{02}} + \sqrt{2} \big( \Ket{11}\Bra{s_{12}} + \Ket{s_{12}}\Bra{22} \big) + {\rm H.c.} \big] +  {\textstyle \frac{1}{2}} \hbar \Omega_{\rm M} \big[ \Ket{s_{02}}\Bra{s_{12}}  +  \Ket{a_{02}}\Bra{a_{12}} \nonumber \\
&& + \sqrt{2} \big( \Ket{00}\Bra{s_{01}} + \Ket{s_{01}}\Bra{11} \big) + {\rm H.c.} \big] + \hbar g \big[
\Ket{s_{01}}\Bra{s_{12}} b^\dagger - \Ket{a_{01}}\Bra{a_{12}} b^\dagger + \sqrt{2} \big( \Ket{00}\Bra{s_{02}} + \Ket{s_{02}}\Bra{22} \big) b^\dagger + {\rm H.c.} \big] \nonumber\\
&& - {\textstyle {\frac{\rm i} {2}}} \hbar \kappa \, b^\dagger b + \hbar \big( \Delta - {\textstyle {\frac{\rm i} {2}}} \Gamma \big) \big[ \Ket{s_{02}}\Bra{s_{02}} +  \Ket{a_{02}}\Bra{a_{02}} + \Ket{s_{12}}\Bra{s_{12}} + \Ket{a_{12}}\Bra{a_{12}} + 2 \, \Ket{22}\Bra{22} \big]  \, .
\end{eqnarray}
\end{widetext}
Similarly, Eq.~(\ref{eqn:Rj}) can be written as
\begin{eqnarray} \label{jump}
{\cal R}_j(\rho) &=& \sum_{i=1,2} R_{ji} \, \rho \, R_{ji}^\dagger
\end{eqnarray}
with the reset operators
\begin{eqnarray} \label{jump3}
R_{01} &\equiv& \sqrt{\Gamma_0} \, \big[ \Ketbra{00}{s_{02}} + \big( \Ketbra{s_{01}}{s_{12}} - \Ketbra{a_{01}}{a_{12}} \big) /\sqrt{2} ~~ \nonumber \\
&& + \Ketbra{s_{02}}{22} \big] \, , \nonumber\\ 
R_{02} &\equiv & \sqrt{\Gamma_0} \, \big[ \Ketbra{00}{a_{02}} + \big( \Ketbra{s_{01}}{a_{12}} - \Ketbra{a_{01}}{s_{12}} \big) /\sqrt{2}  \nonumber \\
&& - \Ketbra{a_{02}}{22} \big]  \, , \nonumber\\
R_{11} &\equiv&  \sqrt{\Gamma_1} \, \big[  \Ketbra{11}{s_{12}} + \big( \Ketbra{s_{01}}{s_{02}} + \Ketbra{a_{01}}{a_{02}} \big) /\sqrt{2}  \nonumber \\
&& + \Ketbra{s_{12}}{22}  \big] \, , \nonumber \\
R_{12} &\equiv&  \sqrt{\Gamma_1} \, \big[ \Ketbra{11}{a_{12}} + \big( \Ketbra{s_{01}}{a_{02}} + \Ketbra{a_{01}}{s_{02}} \big) /\sqrt{2} \nonumber \\
&& - \Ketbra{a_{12}}{22}  \big] \, . 
\end{eqnarray}
That this is indeed the case can be checked by comparing Eqs.~(\ref{jump}) and (\ref{jump3}) with Eq.~(\ref{eqn:Rj}). 

In the following we write the state of the system under the condition of no photon emission, as defined in Eq.~(\ref{eqn:toyEvolution}), as
\begin{eqnarray}
|\psi^0 (t) \rangle &=& \sum_{j,k=0}^2 \sum_{n=0}^\infty \alpha_{jk,n} \, |a_{jk},n \rangle + \sigma_{jk,n} \, |s_{jk},n \rangle \nonumber \\
&& + \sum_{j=0}^2 \sum_{n=0}^\infty \xi_{jj,n} \, |{jj},n \rangle \, .
\end{eqnarray}
Assuming a relatively large detuning $\Delta$, namely as in Eq.~(\ref{eqn:conditions}), the excited atomic states can be eliminated adiabatically. To do so, we use the Schr\"odinger equation and set the derivatives of all coefficients with $j=2$ or $k=2$ equal to zero.  Doing so we find that 
\begin{eqnarray} \label{correct}
\alpha_{02,n} &=& - \frac{\Omega_{\rm L}}{2 \Delta} \, \alpha_{01,n} \, , ~~ \nonumber \\
\alpha_{12,n} &=& \frac{\sqrt{n+1} g}{\Delta} \, \alpha_{01,n+1} \, , ~~ \nonumber \\
\sigma_{02,n} &=& - \frac{\Omega_{\rm L}}{2 \Delta} \, \sigma_{01,n} - \frac{\sqrt{2(n+1)} g}{\Delta} \, \xi_{00,n+1} \, , \nonumber \\
\sigma_{12,n} &=& - \frac{\Omega_{\rm L}}{\sqrt{2} \Delta} \, \xi_{11,n} - \frac{\sqrt{n+1} g}{\Delta} \, \sigma_{01,n+1} \, , \nonumber \\
\xi_{22,n} &=& 0 \, ,
\end{eqnarray}
up to first order in $1/\Delta$. Furthermore, we find that the no-photon time evolution of the system is effectively given by
\begin{eqnarray} \label{cond2}
H_{\rm cond} &=& {\textstyle \frac{1 }{ \sqrt{2}}} \hbar \Omega_{\rm M} \big[ \, |00 \rangle \langle s_{01}| + |s_{01} \rangle \langle 11| + {\rm H.c.} \, \big] \nonumber \\ 
&& + \hbar g_{\rm eff} \big[ \, |00 \rangle \langle s_{01}| b^\dagger+ |s_{01} \rangle \langle 11| b^\dagger + {\rm H.c.} \, \big] \nonumber \\ 
&& + \hbar \big( \Delta_{\rm C} b^\dagger b - \Delta_{\rm L} \big) \big[ \, |00 \rangle \langle 00| - |11 \rangle \langle 11| \, \big] \nonumber \\ 
&& - {\textstyle {\frac{\rm i} { 2}}} \hbar \Gamma_{\rm eff} \big[ \, |a_{01} \rangle \langle a_{01}| +  |s_{01} \rangle \langle s_{01}| + 2 \, |11 \rangle \langle 11| \, \big] \nonumber \\
&& + \hbar \big( \Delta_{\rm C} -  {\textstyle {\frac{\rm i} { 2}}} \kappa \big) \, b^\dagger b
\end{eqnarray}
with the detunings
\begin{eqnarray} \label{cond3}
\Delta_{\rm L} \equiv - \frac{\Omega_{\rm L}^{2}}{4 \Delta} \, , ~~ 
\Delta_{\rm C} \equiv - \frac{g^{2}}{\Delta} 
\end{eqnarray}
and the effective rates
\begin{eqnarray}
g_{\rm eff} \equiv - \frac{\Omega_{\rm L} g}{\sqrt{2} \Delta}  \, , ~~ 
\Gamma_{\rm eff} \equiv \frac{\Omega_{\rm L}^2 \Gamma}{4 \Delta^2} \, .
\end{eqnarray} 
In the derivation of Eq.~(\ref{cond2}), we neglected a constant term in the Hamiltonian with no consequences other than introducing an overall phase shift.

From Eqs.~(\ref{eqn:conditions}), (\ref{cond2}) and (\ref{cond3}), one can see that the evolution of the states with {\em no} photon in the cavity takes place on a time scale much longer than $1/\kappa$. In contrast to this, the time evolution of the states with one or more cavity photons takes place on a time scale proportional to $1/\kappa$. The conditional Hamiltonian (\ref{cond2}) can therefore be simplified further by adiabatically eliminating the states with $n \ge 1$. Doing so, we find that
\begin{eqnarray} \label{correct2}
\xi_{00,1} &=& \frac{\sqrt{2} {\rm i} \Omega_{\rm L} g}{\kappa \Delta} \, \sigma_{01,0} \, , \nonumber \\
\sigma_{01,1} &=& \frac{\sqrt{2} {\rm i} \Omega_{\rm L} g}{\kappa \Delta} \, \xi_{11,0} \, ,
\end{eqnarray}
while all other coefficients with $n \ge 1$ equal zero up to first order in $1/(\kappa \Delta)$. Consequently, Eq.~(\ref{cond2}) becomes
\begin{eqnarray} \label{cond4}
H_{\rm cond} &=& {\textstyle \frac{1 }{ \sqrt{2}}} \hbar \Omega_{\rm M} \big[ \, |00 \rangle \langle s_{01}| + |s_{01} \rangle \langle 11| + {\rm H.c.} \, \big] \nonumber \\ 
&& - \hbar \Delta_{\rm L} \big[ \, |00 \rangle \langle 00| - |11 \rangle \langle 11| \, \big] \nonumber \\
&& - {\textstyle {\frac{\rm i} { 2}}} \hbar \Gamma_{\rm eff} \big[ \, |a_{01} \rangle \langle a_{01}| +  |s_{01} \rangle \langle s_{01}| + 2 \, |11 \rangle \langle 11| \, \big] \nonumber \\
&& - {\textstyle \frac{{\rm i} }{ 2}} \hbar \kappa_{\rm eff} \big[ \, |s_{01} \rangle \langle s_{01}| + |11 \rangle \langle 11| \, \big] 
\end{eqnarray}
with
\begin{eqnarray}
\kappa_{\rm eff} & \equiv & \frac{2 \Omega_{\rm L}^2 g^2}{ \kappa \Delta^2} \, .
\end{eqnarray}
Again, we neglect an overall level shift with no physical consequences. We also neglect detunings of the order $1/\Delta^2$, since these are small compared to other level shifts in $H_{\rm cond}$. We keep all non-Hermitian terms up to the order $1/\Delta^2$. The rate $\Gamma_{\rm eff}$ describes photon emissions from the atoms, while $\kappa_{\rm eff}$ takes the possible leakage of photons through the cavity mirrors into account. Both decay rates scale as $\Omega_{\rm L}^2/\Delta^2$ and are much smaller than $\Gamma$ and $\kappa$.

From Eqs.~(\ref{correct}) and (\ref{correct2}) we see that there is only a small amount of population in the excited states. A photon emission from the atoms can occur when there is population in the excited state $|2 \>$. Eq.~(\ref{correct}) shows that this applies when the atoms are in $|a_{01} \>$, $|s_{01} \>$ or $|11 \>$. The reset operators in Eq.~(\ref{jump3}) change the states $|a_{01} \>$ and $|s_{01} \>$ into $|00 \>$ or transfer $|11 \>$ into $|s_{01} \>$ or $|a_{01} \>$ in case of an emission via the 2--0 transition. Combining Eq.~(\ref{jump3}) with Eq.~(\ref{correct}) and using the notation
\begin{eqnarray}
\Gamma_{{\rm eff};j} \equiv \frac{\Gamma_j \Gamma_{\rm eff} }{ \Gamma} \, ,
\end{eqnarray} 
we obtain 
\begin{eqnarray} \label{RR01}
R_{01} &=& - \sqrt{\Gamma_{{\rm eff}; 0}} \, \big[ \, |00 \> \<s_{01}| + |s_{01} \> \<11| \, \big] \, , \nonumber \\
R_{02} &=& - \sqrt{\Gamma_{{\rm eff}; 0}} \, \big[ \, |00 \> \<a_{01}| - |a_{01} \> \<11| \, \big] \, .
\end{eqnarray}
Analogously, a photon emission via the 2--1 transition is described by  
\begin{eqnarray} \label{RR11}
R_{11} &=& - \sqrt{ \frac{\Gamma_{{\rm eff};1}}{2}} \, \big[ \, |a_{01} \> \<a_{01}| + |s_{01} \> \<s_{01}| + 2 \, |11 \> \< 11| \, \big] \, , \nonumber \\
R_{12} &=& - \sqrt{ \frac{ \Gamma_{{\rm eff};1}}{2}} \, \big[ \, |s_{01} \> \<a_{01}| + |a_{01} \> \<s_{01}| \, \big] \, . ~~
\end{eqnarray}
Eq.~(\ref{correct2}) shows that the leakage of a photon through the cavity mirrors can occur when the atoms are in $|s_{01} \>$ or $|11 \>$. Such an emission transfers the atoms into the states $|00 \>$ and $|s_{01} \>$. Taking this into account and combinig Eq.~(\ref{eqn:Rcav}) with Eq.~(\ref{correct2}) we obtain the effective cavity leakage reset operator
\begin{eqnarray} \label{RRcav}
R_{\rm C} &=& {\rm i} \sqrt{\kappa_{\rm eff}} \, \big[ \, |00 \> \<s_{01}| +  |s_{01} \> \<11| \, \big] \, . 
\end{eqnarray}
Similarly to Section \ref{IIA}, the normalisation of the above reset operators $R_{\rm i}$ has been chosen such that 
\begin{eqnarray} \label{eqn:wcav2}
w_{\rm i} (\psi ) &=& \| \, R_{\rm i} \, |\psi \> \, \|^2 
\end{eqnarray}
is the probability density for the respective decay. Here the index ${\rm i}$ stands for $01$, $02$, $11$, $12$ and ${\rm C}$. 

\subsection{Macroscopic light and dark periods}

As we have seen in Section \ref{sec:IIB}, it is crucial for the occurrence of macroscopic light and dark periods in the fluorescence of a single quantum system, that the system possesses a so-called {\em dark state}. Fig.~\ref{fig:toy model}(b) shows an effective level scheme of the atom-cavity system illustrating the effect of the conditional Hamiltonian (\ref{cond4}) and the reset operators (\ref{RR01})-(\ref{RRcav}). A comparison with the toy model in Fig.~\ref{fig:toy model}(a) suggests that the maximally entangled state $|a_{01} \rangle$ in Eq.~(\ref{a01}) is the only dark state of the atom-cavity system when
\begin{eqnarray} \label{crucial}
\Gamma_{\rm eff} &\ll & \kappa_{\rm eff} \, .
\end{eqnarray}
Under this condition, spontaneous emissions from $|a_{01} \>$ are rare. Moreover, this state does not excite photons in the cavity mode (c.f.~Eq.~(\ref{correct2})), which could cause leakage of a photon through the cavity mirrors. In principle, fulfilling condition (\ref{crucial}) requires $8C \gg 1$. However, as we see below, it holds well enough for our purposes, even when $C$ approaches one. 

When the states $|00 \>$, $|s_{01} \>$ and $|11 \>$ are populated, the laser field with Rabi frequency $\Omega_{\rm L}$ combined with the atom-cavity coupling characterised by $g$ results in the effective coupling constant $g_{\rm eff}$. Population in the states $|s_{01} \rangle$ and $|11 \rangle$ can therefore result in the leakage of a photon through the cavity mirrors with the spontaneous decay rate $\kappa_{\rm eff}$ (c.f.~Fig.~\ref{fig:toy model}(b)). Compared to this, spontaneous emission from the atoms with the decay rate $\Gamma_{\rm eff}$ is almost negligible. The parameter regime (\ref{eqn:conditions}) is analogous to the parameter regime outlined in Eq.~(\ref{eqn:toy_conditions}), if we identify $\Gamma_{\rm L}$ with cavity decay and $\Gamma_{\rm D}$ with spontaneous emission from the atoms. Thus the time evolution of the atom-cavity system is almost identical to that of the toy model introduced in Section \ref{sec:QJ}. 

\begin{figure}
\begin{minipage}{\columnwidth}
\begin{center}
\resizebox{\columnwidth}{!}{\rotatebox{0}{\includegraphics{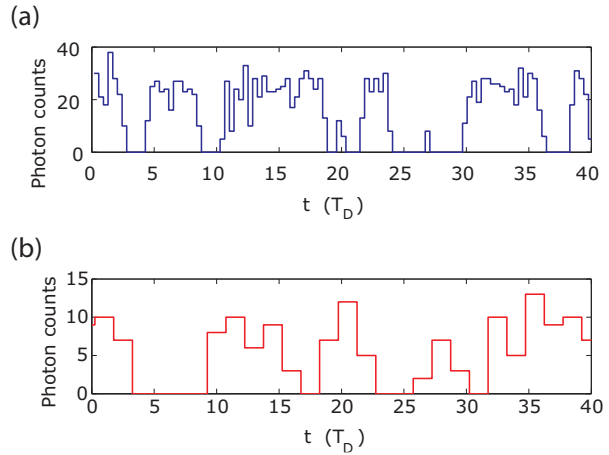}}}
\end{center}
\vspace*{-0.5cm}
\caption{(Colour online) Possible trajectories of the atom-cavity system excibiting macroscopic quantum jumps. The figures have been obtained from quantum jump simulations using Eqs.~(\ref{eqn:Hcond1}), (\ref{eqn:Rj}) and (\ref{eqn:Rcav}) and with  $T_{\rm D}$ as in Eq.~(\ref{eqn:Tdark}). Shown is the number of photon counts within their respective time intervals of length (a) $\Delta t = 0.38 \, T_{\rm D}$ and (b) $\Delta t = 1.5 \, T_{\rm D}$ for maximum detector efficiency ($\eta = 1$). Moreover, $\Delta = 50 \, g$, $\Omega_{\rm L} = \kappa = g$, $\Omega_{\rm M} = 0.05 \, g$, and $\Gamma_0 = \Gamma_1$. In (a) $\Gamma = 0.1 \, g$ and hence $C=10$, while $\Gamma = g$ giving $C=1$ in (b).}
\label{fig:EPRjumps}
\end{minipage}
\end{figure}

We therefore expect to see light and dark periods in the leakage of photons through the cavity mirrors. Within a dark period, the atoms are prepared in the maximally entangled state $|a_{01} \>$. Within a light period, the time evolution of the atom-cavity system remains restricted onto the states $|00 \>$, $|s_{01} \>$ and $|a_{01} \>$. This is confirmed by numerical simulations. Fig.~\ref{fig:EPRjumps} shows possible trajectories of the atom-cavity system based on a quantum jump simulation using the conditional Hamiltonian (\ref{eqn:Hcond1}) and the reset operators (\ref{eqn:Rj}) and (\ref{eqn:Rcav}). Even when the single-atom cooperativity parameter $C$ is as low as one, one can clearly distinguish macroscopic light and dark periods.

\subsection{Characteristic time scales} \label{final}

We now proceed as in Section \ref{sec:ts1} and calculate the characteristic time scales of the atom-cavity system. The mean length of a light period depends primarily on the state of the system within a light period. The steady state $\rho_{\rm ss}$ of the light subspace can be calculated by setting $\dot \rho$ in Eq.~(\ref{eqn:master2}) equal to zero and assuming $\Gamma = 0$. Using Eqs.~(\ref{cond4}) and (\ref{RRcav}) and the notation
\begin{eqnarray} \label{eqn:EPRx}
y \, \equiv \, \frac{\Delta_{\rm L}}{\Omega_{\rm M}} \, = \, - \frac{\Omega_{\rm L}^2}{4 \Delta \Omega_{\rm M}} \, ,
\end{eqnarray}
we obtain
\begin{eqnarray} \label{eqn:ssPops}
\langle 00| \, \rho_{\rm ss} \, |00 \rangle &=& \frac {1+4y^2}{3 + 4 y^2} \, , ~~  \nonumber \\
\langle s_{01}| \, \rho_{\rm ss} \, |s_{01} \rangle &=& \frac{1+ 8 y^2}{3 + 16 y^2 + 16 y^4} \, , ~ \nonumber \\
\langle 11| \, \rho_{\rm ss} \, |11 \rangle &=& \frac{1}{3 + 16 y^2 + 16 y^4} \, .
\end{eqnarray}
Combining these populations with Eq.~(\ref{correct2}), we can now calculate the mean number of photons in the cavity mode. It equals 
\begin{eqnarray}
\langle n \rangle &=& \frac{2 \Omega_{\rm L}^2 g^2}{\kappa^2 \Delta^2} \, \big[ \, \langle s_{01}| \, \rho_{\rm ss} \, |s_{01} \> + \langle 11| \, \rho_{\rm ss} \, |11 \rangle \, \big] \, .
\end{eqnarray}
Since $T_{\rm C}  = 1/\kappa \Braket{n}$, we find that
\begin{equation} \label{eqn:Tcav}
T_{\rm C}  =  \big( 3+4 y^2\big) \cdot \frac{\kappa \Delta^2 }{4 \Omega_{\rm L}^2 g^2} 
\end{equation}
is the mean time between two photon emissions within a light period. 

The small amount of population in excited atomic states occasionally leads to an atomic decay. Proceeding as in Section \ref{sec:ts1} and using Eqs.~(\ref{RR01})-(\ref{eqn:wcav2}), we find that the probability density for this to result in a transition from a light period into a dark period equals $ \Gamma_{\rm eff;0} \, \<11 | \, \rho_{\rm ss} \, |11 \> + \frac{1}{2} \Gamma_{\rm eff;1} \, \<s_{01} | \, \rho_{\rm ss} \, |s_{01} \>$. The inverse of this rate, namely
\begin{equation} \label{eqn:Tlight}
T_{\rm L} =  \frac{3 + 16 y^2 + 16y^4 }{2 \Gamma_0 + (1 + 8 y^2) \Gamma_1} \cdot \frac{8 \Delta^2 }{ \Omega_{\rm L}^2} \, ,
\end{equation}
gives us the mean length of a light period. Analogously, we find that the probability density for a transition from a dark into a light period equals $\Gamma_{\rm eff;0} + \frac{1}{2} \Gamma_{\rm eff;1}$. Hence
\begin{equation} \label{eqn:Tdark}
T_{\rm D} = \frac{1 }{2 \Gamma_0 + \Gamma_1} \cdot  \frac{8 \Delta^2 }{ \Omega_{\rm L}^2} 
\end{equation}
is the mean length of a dark period. 

\begin{figure}[t]
\begin{minipage}{\columnwidth}
\begin{center}
\resizebox{\columnwidth}{!}{\rotatebox{0}{\includegraphics {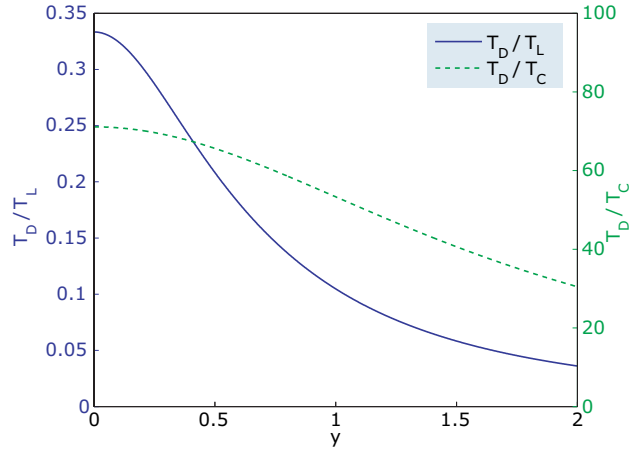}}}
\end{center}
\vspace*{-0.5cm}
\caption{(Colour online) Comparison of the mean length of a dark period $T_{\rm D}$ with the mean length of a light period $T_{\rm L}$ and the mean time $T_{\rm C}$ between cavity photon emissions within a light period as a function of $y$ (c.f.~Eq.~(\ref{eqn:EPRx})) for the same experimental parameters as in Fig.~\ref{fig:EPRjumps}(a).} \label{fig:EPRratios}
\end{minipage}
\end{figure}

In order to ensure the frequent occurrence of dark periods, it is important that $T_{\rm D}$ is not orders of magnitude smaller than $T_{\rm L}$. In addition, it is only easy to distinguish a dark period from a light period, when $T_{\rm D}$ is much larger than $T_{\rm C}$. In analogy to Eq.~(\ref{ratios}), we now find that 
\begin{eqnarray} \label{last}
\frac{T_{\rm D}}{T_{\rm L}} &=& \frac{2 \Gamma_0 + (1 + 8 y^2) \Gamma_1}{(3 + 16 y^2 + 16y^4)(2 \Gamma_0 + \Gamma_1)} \, , \nonumber \\
\frac{T_{\rm D}}{T_{\rm C}} &=& \frac{32 \, g^2}{(3 + 4 \, y^2) (2 \Gamma_0 + \Gamma_1) \, \kappa} \, .
\end{eqnarray}  
Fig.~\ref{fig:EPRratios} shows $T_{\rm D}/T_{\rm L}$ and $T_{\rm D}/T_{\rm C}$ as functions of $y$. To see a clear signature of macroscopic quantum jumps in the fluorescence from the atoms we require $y$ to be close to or smaller than one. This is in good agreement with the parameter regime assumed in Eq.~(\ref{eqn:conditions}). Particularly for the {\em optimal} parameter regime $y \ll1$ and the special case of $\Gamma_0 = \Gamma_1 = \frac{1}{2} \Gamma$, Eq.~(\ref{last}) can be simplified further and reveals that 
\begin{eqnarray} \label{lastlast}
\frac{T_{\rm D}}{T_{\rm L}} = \frac{1}{3}  ~~ {\rm and} ~~ 
\frac{T_{\rm D}}{T_{\rm C}} = \frac{64 C}{9} \, . 
\end{eqnarray}
In this case, light periods are on average three times as long as dark periods and dark periods are about $7C$ times longer than the average time between photons within a light period. Since $\kappa_{\rm eff} / \Gamma_{\rm eff} = 8C$,  Eq.~(\ref{crucial}) is indeed necessary to ensure very large ratios of $T_{\rm D}/T_{\rm C}$. 

\section{Entangled pair generation} \label{entangle}

In the previous section, we have seen that the laser driven atom-cavity system shown in Fig.~\ref{fig:setup} exhibits macroscopic light and dark periods in its fluorescence through the cavity mirrors. The time evolution of the system thereby remains mainly restricted onto atomic ground states. Whenever the fluorescence stops, the two atoms in the cavity are shelved in the maximally entangled state (\ref{a01}). The setup is therefore well suited for the preparation of maximally entangled atom pairs. The completion of the state preparation requires nothing more than turning off the applied laser fields within a dark period. Then the time evolution of the system stops and the atoms remain in $|a_{01} \>$. 

One factor that reduces the fidelity of the final state is the presence of a small amount of population in the excited state $|a_{02} \>$. From Eq.~(\ref{correct}) we see that this population equals
\begin{eqnarray}
|\alpha_{02,0} |^2 &=& \frac{\Omega_{\rm L}^2}{4 \Delta^2} 
\end{eqnarray}
within a dark period. However, for relatively large detunings $\Delta$, such as $\Delta \ge 50 \, \Omega_{\rm L}$, the population in $|a_{02} \>$ is smaller than $10^{-4}$. In general, the corresponding correction to the fidelity of the final state is very small. An error which reduces the fidelity of the final state more significantly, is the possibility to overlook the onset of a light period. The system might decay via an atomic decay into $|00 \>$ or $|s_{01} \>$ when in $|a_{01} \>$. However, this might not yet have resulted in the leakage of a photon through the cavity mirrors and remains undetected. In the following, we calculate the fidelity of the prepared state as a function of the system parameters. We first consider the case of ideal photon detectors before taking finite photon detector efficiencies $\eta < 1$ into account.

\subsection{The Markovian behaviour of the system}

The discussion in Section \ref{entangle1} shows that the time evolution of the system consists mainly of random jumps between periods of no fluorescence into periods of intense fluorescence and vice versa, as illustrated by Fig.~\ref{fig:EPRjumps}. Within a light period, the state of the atom-cavity system is given by the steady state with constant amounts of population in $|00 \>$, $|s_{01} \>$ and $|11 \>$. Within a dark period the atoms are in the state $|a_{01} \>$. The probability density for the occurrence of macroscopic quantum jumps depends therefore only on the current state of the system but not on its evolution in the past. This is typical for {\em Markov processes}, which have been studied in great detail in the literature \cite{Markov}. The analysis of the system is therefore relatively straightforward. 

In the following, we denote the rate with which the system changes from a light period into a dark period by $\gamma_{\rm L}$. Analogously, $\gamma_{\rm D}$ is the probability density for a transition from a dark period into a light period. Moreover, $\gamma_{\rm C}$ is the probability density for the leakage of a photon through the cavity mirrors within a light period. Due to the Markovian behavior of the system, these rates can easily be related to $T_{\rm L}$, $T_{\rm D}$, and $T_{\rm C}$ and we have 
\begin{eqnarray} \label{lostcause}
\gamma_{\rm L} = \frac{1}{T_{\rm L}} \, , ~~
\gamma_{\rm D} = \frac{1}{T_{\rm D}} \, ,  ~~ {\rm and} ~~
\gamma_{\rm C} = \frac{1}{T_{\rm C}} \, .
\end{eqnarray}
These three crucial rates depend on the system parameters and can be obtained directly from
Eqs.~(\ref{eqn:Tcav}), (\ref{eqn:Tlight}) and (\ref{eqn:Tdark}).

The Markovian nature of the time evolution of the system also implies that the probability for remaining in a dark period for a time interval $(0,t)$ equals
\begin{equation} \label{eqn:Pd}
P_{\rm cont \, D}(t) = {\rm e}^{-\gamma_{\rm D} t} \, ,
\end{equation}
given that the system is in its dark state at $t=0$. Moreover,
\begin{equation} \label{eqn:Pl}
P_{\rm cont \, L}(t) = {\rm e}^{-\gamma_{\rm L} t} 
\end{equation}
is the probability for remaining within a light period for a time $t$, given that the system is within the light subspace at $t=0$. Finally, we remark that
\begin{eqnarray} \label{eqn:P0}
P_{\rm 0 \land cont \, L}(t) &= & {\rm e}^{-\gamma_{\rm L} t} \, {\rm e}^{-\gamma_{\rm C} t} =  {\rm e}^{-(\gamma_{\rm L} + \gamma_{\rm C}) t}
\end{eqnarray}
is the probability of remaining in a light period without emitting a single cavity photon for a time $t$, given that the system is initially in a light period. Other probabilities characterising the time evolution of the system can be calculated in an analogous way.

\subsection{Fidelity of the prepared state for unit photon detector efficiency}

Here we are particularly interested in the case, where the system is in a light period and a photon has just been detected at time $t=0$. We then ask the question, what is the probability $P_{\rm 0 \land D}(t)$ of observing no photon for a time $t$ {\em and} to find the system in a dark state at $t$. Analogously, $P_{\rm 0 \land L}(t)$ is the probability of finding no photon in $(0,t)$ {\em and} finding the system in the light subspace at $t$. The fidelity of the state prepared after the detection of no photon in $(0,t)$ can then be written as
\begin{eqnarray} \label{eqn:F(t)}
F(t) &=& \frac{P_{\rm 0 \land D}(t)}{P_{\rm 0 \land D}(t) + P_{\rm 0 \land L}(t)} \, .
\end{eqnarray}
Let us now calculate the two probabilities in this equation.

To do so, we denote the probability of finding the system in a dark period at time $t$ after undergoing $n$ light periods without a single photon emission by $Q_n(t)$. Then 
\begin{eqnarray} \label{eqn:0+dark}
P_{\rm 0 \land D}(t) &=& \sum_{n=1}^\infty Q_n(t) \, .
\end{eqnarray}
Suppose $t_1$ denotes the time when the system switches from the initial light period into a dark period, which happens with probability density $\gamma_{\rm L}$. Then one can show by using Eqs.~(\ref{eqn:Pd}) and (\ref{eqn:P0}) that
\begin{eqnarray} \label{eqn:Q1}
Q_1(t) &=& \gamma_{\rm L} \, \int_{0}^{t} {\rm d}t_1 \, {\rm e}^{-\gamma_{\rm D} (t - t_1)} \, {\rm e}^{ - (\gamma_{\rm L} + \gamma_{\rm C}) t_1} \, . ~~
\end{eqnarray}
Similarly, one can show that 
\begin{eqnarray}
Q_2(t) &=& \gamma_{\rm L}^2 \gamma_{\rm D} \, \int_{0}^{t} {\rm d}t_3 \int_0^{t_3} {\rm d}t_2 \int_0^{t_2} {\rm d}t_1 \nonumber \\
&& {\rm e}^{-\gamma_{\rm D} (t - t_3)} \, {\rm e}^{ - (\gamma_{\rm L} + \gamma_{\rm C}) (t_3-t_2)} \nonumber \\
&& \times {\rm e}^{-\gamma_{\rm D} (t_2 - t_1)} \, {\rm e}^{- (\gamma_{\rm L} + \gamma_{\rm C}) t_1} \, . 
\end{eqnarray}
Here $t_1$ and $t_3$ denote transitions from a light into a dark period and $t_2$ marks a transition from a dark to a light period. More generally,
\begin{eqnarray} \label{eqn:Qn}
Q_n (t) &=& \gamma_{\rm L}^n \, \gamma_{\rm D}^{n-1} \, \int_{0}^{t} {\rm d}t_{2n-1} \int_0^{t_{2n-1}} {\rm d}t_{2n-2} \ldots \int_{0}^{t_2} {\rm d}t_1 \nonumber \\
&& {\rm e}^{-\gamma_{\rm D}(t-t_{2n -1})} \, {\rm e}^{-(\gamma_{\rm L}+\gamma_{\rm C})(t_{2n-1} -t_{2n-2})} \nonumber \\
&& \times \ldots \times {\rm e}^{-\gamma_{\rm D}(t_2-t_1)} \, {\rm e}^{-(\gamma_{\rm L} + \gamma_{\rm C}) t_1} \, .
\end{eqnarray}
In order to evaluate these nested integrals up to infinite depth and to calculate $P_{\rm 0 \land D}(t)$ in Eq.~(\ref{eqn:0+dark}), we note that they correspond to infinite depth convolution integrals. We therefore make use of the Laplace transform ${\cal L}$, which is similar to the Fourier transform of a function but has the properties \cite{Laplace}
\begin{eqnarray}
&& {\cal L}\left( f(t) + g(t) \right) = {\cal L} (f(t)) + {\cal L}(g(t)) \, , \nonumber \\ 
&& {\cal L}\left( \int_0^t  {\rm d}\tau \, f(t-\tau)g(\tau) \right) = {\cal L} (f(t)) \, {\cal L}(g(t)) \, . ~~~~~
\end{eqnarray}
Using these two rules, we find that
\begin{eqnarray}
{\cal L}( P_{\rm 0 \land D} (t) )&=& \sum_{n=1}^\infty \frac{\gamma_{\rm L}^{n} \gamma_{\rm D}^{n-1}}{(s+\gamma_{\rm D})^n (s + \gamma_{\rm L} + \gamma_{\rm C})^n} \, . \nonumber \\ 
\end{eqnarray}
Evaluating this expression, we obtain
\begin{eqnarray}
{\cal L} \left( P_{\rm 0 \land D}(t) \right) &=& 
{\frac {{ \gamma_{\rm L}}}{{s}^{2} + \left( { \gamma_{\rm L}}+{ \gamma_{\rm C}}+{ \gamma_{\rm D}} \right) s
+{ \gamma_{\rm D}}\,{ \gamma_{\rm C}}}} \, . \nonumber \\ 
\end{eqnarray}
Hence, the probability of being in a dark period at $t$ without any cavity photon emissions in $(0,t)$
\begin{eqnarray} \label{eqn:P0D}
P_{\rm 0 \land D}(t) &=& {\frac {{ \gamma_{\rm L}}}{A}} \, \sinh \left( A t \right) \,{{\rm e}^{- ( { \gamma_{\rm L}}+{ \gamma_{\rm C}}+{ \gamma_{\rm D}} ) t / 2}} \, ,
\end{eqnarray}
which is used later to determine the fidelity $F(t)$ in Eq.~(\ref{eqn:F(t)}).

\begin{figure}
\begin{minipage}{\columnwidth}
\begin{center}
\resizebox{\columnwidth}{!}{\rotatebox{0}{\includegraphics{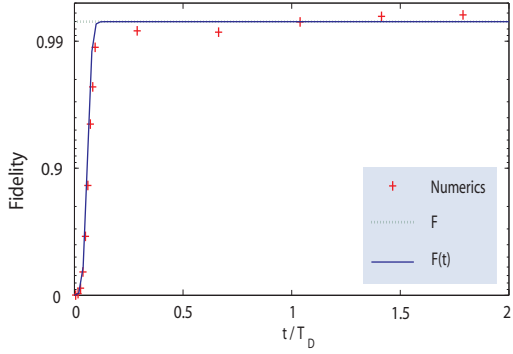}}}
\end{center}
\vspace*{-0.5cm}
\caption{(Colour online) The fidelity $F(t)$ of the state prepared after the detection of no photon for a time $t$ obtained from an evaluation of Eq.~(\ref{eqn:Ffinal}) (solid line), of Eq.~(\ref{lastlastlastlast}) (dashed line), and from a quantum jump simulation (marked by crosses). The system parameters are $\eta=1$, $\Delta = 50 \, g$, $\Omega_{\rm L} = \kappa = g$, $\Omega_{\rm M} = \Gamma = 0.05 \, g$, and $\Gamma_0 = \Gamma_1$. Hence $C=20$.}
\label{fig:FidelityEta1}
\end{minipage}
\end{figure}

In an analogous way, we now calculate the probability $P_{\rm 0 \land L}(t)$ in Eq.~(\ref{eqn:F(t)}). It is given by  
\begin{eqnarray} \label{eqn:0+light}
P_{\rm 0 \land L}(t) &=& \sum_{n=1}^\infty R_n(t) 
\end{eqnarray}
when $R_n(t)$ denotes the probability of finding the system in a light period after experiencing $n$ light periods without any cavity photon emissions. Proceeding as above and assuming a photon emission within a light period at $t=0$, we find that $R_1(t) = P_{\rm 0 \land cont \, L}(t)$ in Eq.~(\ref{eqn:P0}). More generally 
\begin{eqnarray} \label{eqn:Rn}
R_n (t) &=& \gamma_{\rm L}^{n-1} \, \gamma_{\rm D}^{n-1} \, \int_{0}^{t} {\rm d}t_{2n-2} \int_0^{t_{2n-3}} {\rm d}t_{2n} \ldots \, \int_{0}^{t_2} {\rm d}t_1 \nonumber \\
&& {\rm e}^{-(\gamma_{\rm L}+\gamma_{\rm C})(t - t_{2n-2})} \, {\rm e}^{-\gamma_{\rm D}(t_{2n-2} -t_{2n - 3})} \nonumber \\
&& \times \ldots \times {\rm e}^{-\gamma_{\rm D}(t_2-t_1)} \, {\rm e}^{-(\gamma_{\rm L} + \gamma_{\rm C}) t_1} \, .
\end{eqnarray}
Applying the Laplace transformation to this function, using Eq.~(\ref{eqn:0+light}) and proceeding as above, we then obtain
\begin{eqnarray}
{\cal L}(P_{\rm 0 \land L} (t)) &=& \sum_{n=1}^\infty \frac{\gamma_{\rm L}^{n-1} \, \gamma_{\rm D}^{n-1}}{(s+\gamma_{\rm D})^{n-1} (s + \gamma_{\rm L} + \gamma_{\rm C})^n} \, . \nonumber \\ 
\end{eqnarray}
Performing the summation, we find 
\begin{eqnarray}
{\cal L}(P_{\rm 0 \land L} (t)) &=& {\frac {s+{ \gamma_{\rm D}}}{{s}^{2}+ \left( { \gamma_{\rm L}}+{ \gamma_{\rm C}}+{ \gamma_{\rm D}}
 \right) s+{ \gamma_{\rm D}}\,{ \gamma_{\rm C}}}} \, . \nonumber \\ 
\end{eqnarray}
The inverse transform of this function is the probability of being in a light period at $t$ without any cavity photon emissions in $(0,t)$. It is given by
\begin{eqnarray}
P_{\rm 0 \land L} (t) &=& \Bigg[ \, \cosh \left( A t \right) - 
\frac{ \gamma_{\rm L} + \gamma_{\rm C} - \gamma_{\rm D}}{2 A} \,  \sinh ( A t ) \, \Bigg] \nonumber \\
&& \times {\rm e}^{- \left( { \gamma_{\rm L}}+{ \gamma_{\rm C}}+{ \gamma_{\rm D}} \right) t /2} \, . 
\end{eqnarray}
Substituting this and Eq.~(\ref{eqn:P0D}) into Eq.~(\ref{eqn:F(t)}), we finally arrive at an expression for the fidelity of the prepared state of the two atoms after turning off the laser fields upon the detection of no cavity photon for a time $t$. It equals
\begin{eqnarray} \label{eqn:Ffinal}
F(t) &=& \frac{2 \gamma _{L} \sinh (A t)}{2 A \cosh (At) - (\gamma_{\rm C} - \gamma_{\rm D} - \gamma_{\rm L}  ) \, \sinh (At)} \nonumber \\ 
\end{eqnarray}
with 
\begin{eqnarray} \label{A}
 A &\equiv & {\textstyle \frac{1}{2}} \, \Big[ \big(\gamma_{\rm C} + \gamma_{\rm D} + \gamma_{\rm L} \big)^2 - 4 \gamma_{\rm C} \gamma_{\rm D} \Big]^{1/2} \, .
\end{eqnarray}
The parameter $A$ is in general found to be close to but also smaller than $\frac{1}{2} \gamma_{\rm C}$, when evaluated for concrete experimental parameters. Fig.~\ref{fig:FidelityEta1} shows a very good agreement between the fidelity in Eq.~(\ref{eqn:Ffinal}) and the fidelity obtained from a quantum jump simulation for the same set of experimental parameters.  

Particularly for the optimal parameter regime $y \ll1$ and the special case of $\Gamma_0 = \Gamma_1 = \frac{1}{2} \Gamma$, we see from Eqs.~(\ref{lastlast}) and (\ref{lostcause}) that
\begin{eqnarray}
\gamma_{\rm L} = \frac{3 \gamma_{\rm C}}{64 C}  ~~ {\rm and} ~~
\gamma_{\rm D} = \frac{9 \gamma_{\rm C}}{64 C} \, .
\end{eqnarray}
Substituting this into Eqs.~(\ref{eqn:Ffinal}) and (\ref{A}), we can now calculate the fidelity of the final state as a function of the single-atom cooperativity parameter $C$ alone. In the limit of large times $t$, this yields an estimate of the achievable fidelity using the above described state preparation scheme. The asymptotic expression $F \equiv \lim_{t \to \infty} F(t)$ that we obtain is
\begin{eqnarray} \label{lastlastlastlast}
F &=& \frac{3}{2 \big[ \big(256 C^2 - 48 C + 9 \big)^{1/2} - 16 C +3 \big]} \, . ~~~
\end{eqnarray}
This agrees very well with the results obtained from quantum jump simulations using Eqs.~(\ref{eqn:Hcond1}), (\ref{eqn:Rj}) and (\ref{eqn:Rcav}), as illustrated in Fig.~\ref{fig:FidelityEta1}. 

The fidelity in Eq.~(\ref{lastlastlastlast}) tends to unity for very large $C$, i.e.~in the so-called strong coupling regime. However, even for $C=1$, we find that the fidelity of the prepared state can be above $0.86$. The proposed entangled state preparation scheme is therefore expected to operate well even in the vicinity of the bad cavity limit $(C=1)$. For example, for $C=10$, we obtain $F > 0.98$. For larger $C$'s, the same high fidelities are achievable even when using imperfect photon detectors.

\subsection{Finite photon detector efficiencies}

The effect of finite photon detector efficiencies $\eta <1$ is to increase the effective time between two detector clicks during a light period. More concretely, $T_{\rm C}$ becomes $T_{\rm C}/\eta$. If we define the cavity photon detection rate as the average number of photons to be {\em detected} per unit time, then 
\begin{eqnarray} \label{eta1}
\gamma_{\rm C} \longrightarrow \eta \,  \gamma_{\rm C} \, .
\end{eqnarray}
However, the mean duration of the light and dark periods, $T_{\rm L}$ and $T_{\rm D}$, are macroscopic signal properties. They are mostly unaffected by changes in $\eta$ and remain the same, as long as $\eta$ is not so small that $T_{\rm C}/ \eta$ becomes comparable to $T_{\rm D}$, in which case no clear macroscopic jump signal would be seen. The rates describing the transition from a light into a dark period and vice versa are the same for all $\eta$.

\begin{figure}
\begin{minipage}{\columnwidth}
\begin{center}
\resizebox{\columnwidth}{!}{\rotatebox{0}{\includegraphics{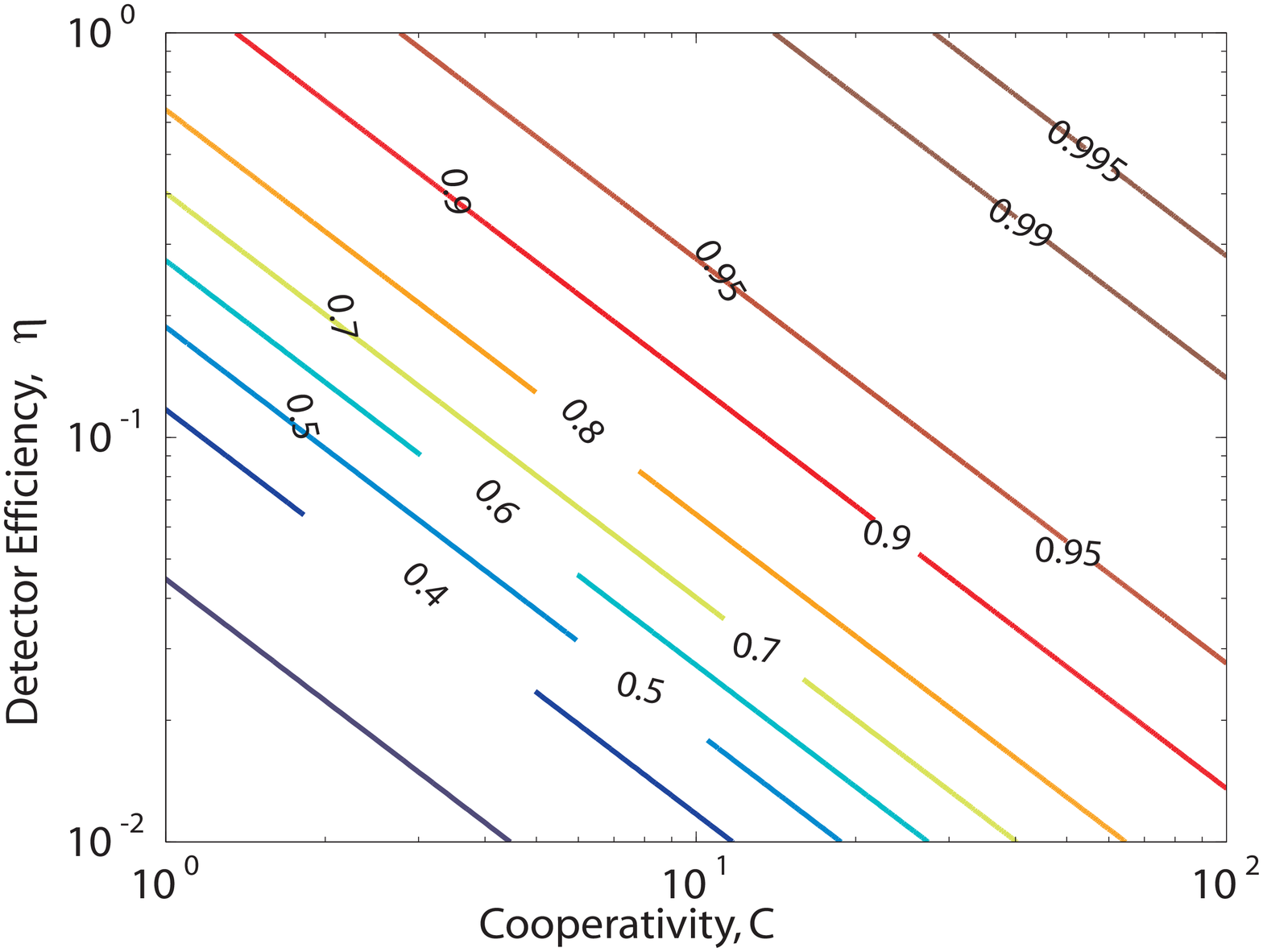}}}
\end{center}
\vspace*{-0.5cm}
\caption{(Colour online) Log-log contour plot of the asymptotic fidelity $F$ in Eq.~(\ref{eqn:FfinalEta}) for different detector efficiencies $\eta$ and for different atom-cavity cooperativity parameters $C$. This fidelity is a good estimate of the achievable precision of the proposed state preparation scheme.}
\label{fig:FidelityContour}
\end{minipage}
\end{figure}

The fidelity of the final state for $\eta < 1$ can now be calculated in the same way as in the previous section, namely by considering a Markov process. The final result is the same but with $\gamma_{\rm C}$ replaced by $\eta \, \gamma_{\rm C}$. The fidelity of the state of the two atoms in the cavity prepared after the {\em detection} of no cavity photon for a time $t$ therefore equals
\begin{eqnarray} \label{lastlastlast}
F(t) &=& \frac{2 \gamma _{L} \sinh (A t)}{2 A \cosh (At) - (\eta \, \gamma_{\rm C} - \gamma_{\rm D} - \gamma_{\rm L}  ) \, \sinh (At)}  \nonumber \\
\end{eqnarray}
with
\begin{eqnarray} \label{lllll}
 A &\equiv & {\textstyle \frac{1}{2}} \, \Big[ \big(\eta \, \gamma_{\rm C} + \gamma_{\rm D} + \gamma_{\rm L} \big)^2 - 4 \eta \, \gamma_{\rm C} \gamma_{\rm D} \Big]^{1/2} \, .
\end{eqnarray}
Again, for the optimal parameter regime $y \ll1$ and the special case of $\Gamma_0 = \Gamma_1 = \frac{1}{2} \Gamma$, Eqs.~(\ref{lastlast}) and (\ref{lostcause}) can be used to calculate the asymptotic limit of this fidelity for large $t$. It is given by 
\begin{eqnarray} \label{eqn:FfinalEta}
F &=& \frac{3}{2 \big[ \big(256 \eta^2 C^2 - 48 \eta C + 9 \big)^{1/2} - 16 \eta C +3 \big]}  \nonumber \\
\end{eqnarray}
and is the same as in Eq.~(\ref{eqn:Ffinal}) but with $C$ replaced by $\eta C$. Achieving fidelities for the entangled state generation above $0.86$ therefore requires $\eta C \ge 1$, as illustrated in Fig.~\ref{fig:FidelityContour}.

\subsection{Robustness against parameter fluctuations}

In deriving the above results we assumed that the atoms both see the same effective cavity coupling strength $g$ and the same laser Rabi frequencies $\Omega_{\rm M}$ and $\Omega_{\rm L}$. This was to ensure the symmetric Hamiltonian (\ref{cond4}) with separate time evolutions of the symmetric and the antisymmetric state space. However, in a realistic setup there will be corrections to this and the Hamiltonian will contain an element of asymmetry, which induces an additional transition between both subspaces. In the following, we analyse the effect of variations of the effective atom-cavity coupling $g_{\rm eff}$ and the frequency $\Omega_{\rm M}$ between both atoms.  

\begin{figure}
\begin{minipage}{\columnwidth}
\begin{center}
\resizebox{\columnwidth}{!}{\rotatebox{0}{\includegraphics{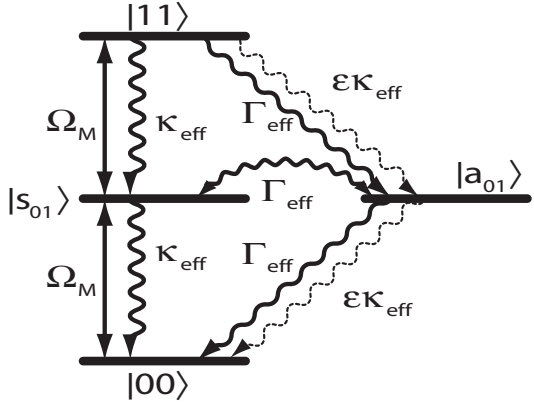}}}
\end{center}
\vspace*{-0.5cm}
\caption{Effective level scheme for the atomic system when $g_{1} \ne g_{2}$. This leads to different effective decay rates $\kappa_{\rm eff}$ which cause additional leakage between the symmetric and antisymmetric subspaces.}
\label{fig:robustlevels}
\end{minipage}
\end{figure}

\begin{figure}
\begin{minipage}{\columnwidth}
\begin{center}
\resizebox{\columnwidth}{!}{\rotatebox{0}{\includegraphics{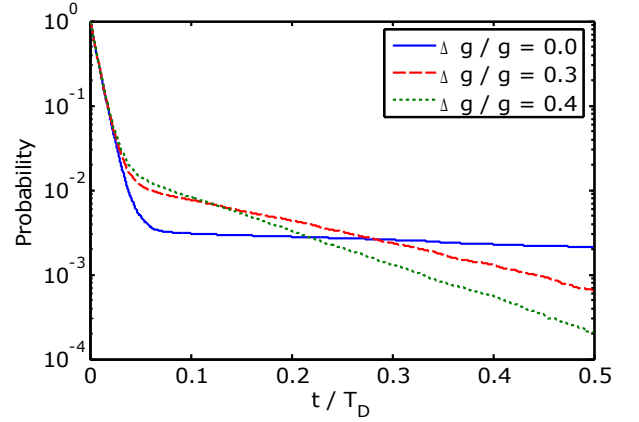}}}
\end{center}
\vspace*{-0.5cm}
\caption{(Colour online) Probability density for the observation of {\em no} photon for a certain time $t/T_{\rm D}$ for different $\Delta g / g$ obtained from a numerical simulation of the time evolution of the systems after averaging over many trajectories. The parameters are the same parameters as in Fig.~\ref{fig:FidelityEta1} and $T_{\rm D}$ is the mean length of a dark period for $\Delta g=0$. Long dark periods become less likely as the difference between $g_1$ and $g_2$ increases. This decrease is  negligible as long as $(\Delta g / g)^2 \ll 1$.}
\label{fig:robustprob}
\end{minipage}
\end{figure}

\begin{figure}
\begin{minipage}{\columnwidth}
\begin{center}
\resizebox{\columnwidth}{!}{\rotatebox{0}{\includegraphics{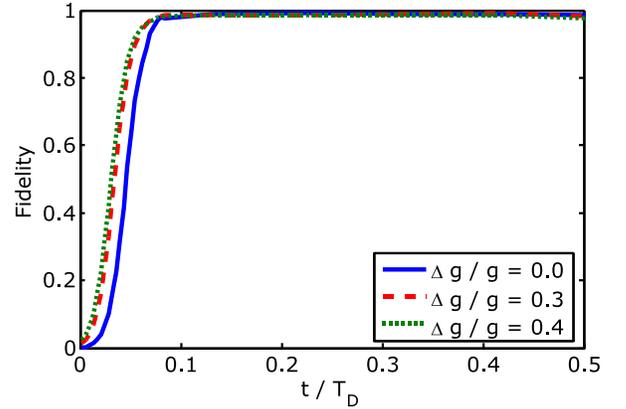}}}
\end{center}
\vspace*{-0.5cm}
\caption{(Colour online) Fidelity of the prepared state as a function of $t/T_{\rm D}$ for the same parameters as in Fig.~\ref{fig:robustprob}. Here $t$ is the time it takes to complete the state preparation, i.e.~the time after which the laser field is turned off upon the detection of no cavity photon since the last event. We see that it is possible to obtain very high fidelities, even when $\Delta g$ is relatively large.}
\label{fig:robustfid}
\end{minipage}
\end{figure}

Suppose, atom 1 and atom 2 experience different cavity coupling constants $g_1$ and $g_2$. Then the anti-symmetric state $|a_{01}\>$ is no longer decoupled from the cavity field but couples to it at a rate proportional to $\Delta g \equiv g_1 - g_2$. This can cause a cavity photon emission and a transition from $|a_{01} \rangle$ to $|00 \rangle$, as illustrated in Fig.~\ref{fig:robustlevels}. The spontaneous decay rate of $|a_{01} \rangle$ equals $\epsilon \cdot \kappa_{\rm eff}$, where $\epsilon$ is proportional to $(\Delta g/g)^2$ with $g \equiv {1 \over 2} (g_1+g_2)$. Fig.~\ref{fig:robustprob} shows the probability to detect no photon for a minimum time $t$ as a function of $\Delta g /g$. We see that longer dark periods become more rare as $(\Delta g/g)^2$ increases. In the case of different the atom-cavity coupling constants,  on average the preparation of the maximally entangled state therefore takes longer. However, the antisymmetric state remains the state with the lowest spontaneous decay in the system, as long as $\Omega_{M}$ is the same for both atoms. The fidelity of the state prepared upon the detection of no photon for a certain time $t$ is hence only minimally affected. As shown in Fig. \ref{fig:robustfid}, it is possible to guaranetee high fidelities even when $\Delta g$ approaches $0.5 \, g$. 

Varying $\Omega_{\rm L}$ has the same effect as varying $g$. Both contribute only to the effective atom-cavity coupling constant $g_{\rm eff}$. Slightly more damaging are variations of the Rabi frequency $\Omega_{M}$ across the atoms, since they have the effect of coherently driving population out of the antisymmetric subspace. This then decreases the time, in which the system remains on average in the antisymmetric state. If $\Omega_{M}$ varies by a moderate amount across the atoms, the dark periods cease to be visible and the fidelity of the prepared state is heavily degraded. However, variations of $\Omega_{\rm M}$ up to a few percent can be tolerated.

\section{Conclusions} \label{sec:conc}

Recently, we predicted the occurrence of macroscopic quantum jumps in the fluorescence of a laser-driven atom-cavity system \cite{MTB-PRL}. When a detector monitors the intensity of the light leaking through the cavity mirrors, it sees long periods of fluorescence randomly interrupted by long periods of no fluorescence, as shown in Fig.~\ref{fig:setup}. Here we show that it is possible to clearly distinguish macroscopic light and dark periods, even in the vicinity of the bad cavity limit, where the single atom-cooperativity $C$ is as low as one (c.f.~Fig.~\ref{fig:EPRjumps}(b)). This is possible, since dissipation plays a crucial role in the generation of the different fluorescence signals. The emission of photons continuously reveals information about the system, thereby restricting its state onto a certain subspace of states. Cavity decay is responsible for the emission of photons within a light period, while spontaneous emission from the atoms is responsible for transitions from one fluorescence period into another. 

In Section \ref{sec:QJ}, we discuss the origin of macroscopic quantum jumps in detail by analysing a simple four-level toy model (c.f.~Fig.~\ref{fig:toy model}(a)). Numerical quantum jump simulations are used to predict the possible trajectories of the system. Afterwards, we calculate its characteristic time scales, such as the mean length of a dark period, analytically. The insight obtained in Section~\ref{sec:QJ} is used in Section~\ref{entangle1} to analyse the more complex laser-driven atom-cavity system. We show that its effective level scheme is essentially equivalent to the level structure of the toy model (c.f.~Fig.~\ref{fig:toy model}(b)). We then calculate the mean length of the light and dark periods of the atom-cavity system as well as the mean time between photon emissions within a light period (c.f.~Eqs.~(\ref{eqn:Tcav})-(\ref{eqn:Tdark})).

The applied interactions are the same for both atoms and only infrequent atomic emission events can change the symmetry of the atomic state. We show that the atoms consequently remain within a symmetric state during a light period. In a dark period, the atoms are shelved in the antisymmetric and maximally entangled ground state (\ref{a01}). The setup shown in Fig.~\ref{fig:setup} can therefore be used for the generation of maximally entangled atom pairs. If the applied laser fields are switched off, when a dark period occurs, then the time evolution of the system stops and the atoms remain entangled.

The result is a state preparation scheme for maximally entangled atom pairs that operates with high fidelities even in the vicinity of the bad cavity limit. In Section \ref{entangle} we calculate the fidelity of the prepared state under the condition of no cavity photon {\em detection} for a time $t$ as a function of the experimental parameters and the photon detector efficiency $\eta$ (c.f.~Eqs.~(\ref{lastlastlast}) and (\ref{lllll})). This fidelity depends predominantly on the product $\eta C$   (c.f.~Eqs.~(\ref{eqn:FfinalEta})). Achieving fidelities above $0.86$ is possible, even when $\eta C = 1$. As a result, the described state preparation scheme opens new perspectives for high-precision quantum computing without the necessity for unrealistically efficient setups. 

We have also seen that the scheme is particularly robust against parameter fluctuations across the atoms due to its postselective nature. Variations of the atom-cavity coupling constants $g_1$ and $g_2$ and very small variations of the laser Rabi frequency $\Omega_{\rm M}$ only result in a decrease in the mean length of the dark period and only minimally affect the fidelity of the final prepared state. If the cooperativity parameter $C$ is relatively large, the mean length of a dark period is relatively long and a decrease of the times without photon emissions due to parameter fluctuations can be easily tolerated. \\ 

\noindent {\em Acknowledgment.} We thank P. L. Knight, C. Sch\"on, and M. Trupke for interesting and stimulating discussions. A. B. acknowledges support from the Royal Society and the GCHQ. This work was supported in part by the EU Integrated Project SCALA, the EU Research and Training Network EMALI and the UK Engineering and Physical Sciences Research Council through the QIP IRC.

\end{document}